\documentstyle[12pt]{article}
\newlength{\bredde}
\def\slash#1{\settowidth{\bredde}{$#1$}\ifmmode\,\raisebox{.15ex}{/}
\hspace*{-\bredde} #1\else$\,\raisebox{.15ex}{/}\hspace*{-\bredde} #1$\fi}
\newcommand{\beq}{\begin{equation}}
\newcommand{\eeq}{\end{equation}}

\def\gtwid{\raise.3ex\hbox{$>$\kern-.75em\lower1ex\hbox{$\sim$}}}
\def\ltwid{\raise.3ex\hbox{$<$\kern-.75em\lower1ex\hbox{$\sim$}}}
\begin{document}
\topmargin -0.8cm
\oddsidemargin -0.8cm
\evensidemargin -0.8cm
\title{\Large{
Schwinger--Dyson BRST symmetry \\
and the Batalin--Vilkovisky \\ Lagrangian
Quantisation of Gauge Theories \\ with Open or Reducible Gauge Algebras}}
\vspace{0.5cm}

\author{{\sc Frank De Jonghe}\thanks{
Aspirant of the N.F.W.O., Belgium.}\\
Instituut voor Theoretische Fysica,K.U.Leuven, Belgium \\
and \\
CERN -- Geneva
}
\maketitle
\vfill
\begin{abstract} In this short note we extend the results of Alfaro and
Damgaard on the origin of antifields to theories with a gauge algebra
that is open or reducible.
\end{abstract}
\vspace{6cm}
\begin{flushleft}
KUL-TF-93/13 \\
CERN--TH-6858/93 \\
hepth@xxx/9304025 \\
April 1993
\end{flushleft}
\thispagestyle{empty}
\newpage

\setcounter{page}{1}

\section{Introduction}
In a recent paper \cite{us}, the antifields of the Lagrangian
quantisation scheme of Batalin and Vilkovisky
\footnote{We will refer to this scheme as BV.
For recent reviews, see \cite{review}} \cite{Batalin}
were unmasked as being the antighosts of the Schwinger--Dyson
BRST symmetry \cite{us0}. The latter is implemented most
transparently using the so-called collective field formalism.
Here it amounts to replacing the fields everywhere by the difference
between the field itself and its collective partner. Of course, this
introduces a new symmetry, the shift symmetry, where both the
field and its collective partner are shifted by an arbitrary
amount. The most general Schwinger--Dyson
equations, defining the
complete quantum theory, can then be obtained as Ward identities
of this BRST shift symmetry. Alfaro and Damgaard showed \cite{us}
that this BRST shift symmetry can be combined with gauge symmetries
originally present in the action. Fixing the collective field to
zero and integrating over the ghost field of the shift symmetry,
it is seen that the antighost of the shift symmetry is to be
identified with the BV antifield.

In \cite{Mijzelf1}, it was shown how using the same collective-field
technique, also the extended BRST--anti-BRST Lagrangian
formalism of \cite{BLT} can be recovered. There, to every
field are associated three antifields, which in the collective-field
approach can be identified with the ghosts and antighosts of
the shift symmetry, and with the collective field itself, which
in this case cannot be integrated out. The Schwinger--Dyson
symmetry has also been implemented in the Hamiltonian quantisation
scheme of \cite{BFV}, and this way it is possible to prove the
equivalence of Hamiltonian and Lagrangian BRST quantisation in
a direct and natural way \cite{Mijzelf2}.

However, all the above developments were only valid for closed,
irreducible algebras. It is our purpose in this short note to show
that the treatment of theories with open and/or reducible
algebras does not present any particular difficulty. In contrast
to the original, very algebraic derivation of BV, a more intuitive
introduction of this formidable scheme becomes possible for
all types of gauge theories known today.

In section 2, we briefly review the results of \cite{us}, in
order to make our subsequent treatment of open and reducible
algebras more accessible. The former are discussed in section
3, the latter in section 4.

\section{Review of the procedure}

Here, we will briefly review the steps followed in \cite{us} to
develop the antifield formalism. We start from
$S_0(\phi)$, the original action. Its gauge invariances are generated by
$R^i_{\alpha}(\phi)$, which satisfy
\beq  \frac{\delta^r S_0}{\delta \phi^i} R^i_{\alpha} \epsilon^{\alpha}
= 0. \eeq
The commutator of two gauge transformations is a linear combination of
gauge transformations (see (\ref{algebra}) below, with $E^{ji}_{\alpha
\beta} = 0 $) for closed algebras.
The na\"{\i}ve BRST operator can be constructed
\begin{eqnarray}  \label{deltaN}
    \delta_N \phi^i & = & R^i_{\alpha} c^{\alpha} \nonumber \\
    \delta_N c^{\alpha} & = & T^{\alpha}_{\beta \gamma} c^
{\gamma} c^{\beta}   \nonumber \\
    \delta_N \bar{c}_{\alpha} & = & b_{\alpha}  \\
    \delta_N b_{\alpha} & = & 0 , \nonumber
\end{eqnarray}
and is nilpotent.

We start by collectively denoting the fields $\phi^i$, $c^
{\alpha}$, $\bar{c}_{\alpha}$ and $b_{\alpha}$ by $Y^{\bar{a}}$.
The na\"{\i}ve BRST transformations of (\ref{deltaN}) are then
summarized in the statement
\beq \delta_N Y^{\bar{a}} = {\cal R}^{\bar{a}} (Y^{\bar{a}}) .
\eeq
We now introduce collective fields ${\cal Y}^{\bar{a}}$. The
ghosts and antighosts of the corresponding Schwinger--Dyson BRST
symmetry, we denote respectively by $c^{\bar{a}}$ and $Y^*_{\bar{a}}$.
Following \cite{us}, we organise the combined na\"{\i}ve BRST
transformations as follows:
\begin{eqnarray} \label{deltaNshift}
  \delta_N Y^{\bar{a}} & = & c^{\bar{a}} \nonumber \\
  \delta_N {\cal Y}^{\bar{a}} & = & c^{\bar{a}} - {\cal R}
      ^{\bar{a}} (Y-{\cal Y}) \nonumber \\
  \delta_N c^{\bar{a}} & = & 0 \\
  \delta_N Y^*_{\bar{a}} & = & B_{\bar{a}} \nonumber \\
  \delta_N B_{\bar{a}} & = & 0 . \nonumber
\end{eqnarray}

Gauge-fixing the collective field to zero and gauge-fixing the original
symmetries is then done by adding $\delta_N \left( Y^*_{\bar{a}}
{\cal Y}^{\bar{a}} + \Psi(Y) \right) = \delta_N X$ to the original
action.
It was then recognised that
\beq S_{gf} = S_0(\phi-\varphi) + \delta_N X =
S_{BV} (Y- {\cal Y},Y^*) + (-1)^{\bar{a}}
B_{\bar{a}} {\cal Y}^{\bar{a}} + Y^*_{\bar{a}} c^{\bar{a}}
+ \frac{\delta^r \Psi}{\delta Y^{\bar{a}}} c^{\bar{a}} .\eeq
The statement that $\delta_N S_{gf} = 0$ then immediately implies
that $S_{BV}$ satisfies the BV classical master equation:
\beq \frac{\delta^r S_{BV}}{\delta Y^{\bar{a}}}\cdot \frac
{\delta^l S_{BV}}{\delta Y^*_{\bar{a}}}  = 0.\eeq
The antighosts act as sources for the BRST transformation of the
original symmetries, so that also the BV boundary conditions are
satisfied.

Of course, when considering the path integral of the theory, the
classical action might have to be supplemented by terms of higher
order in $\hbar$, because the possible non-invariance of the measure
under BRST transformations
may spoil the usual derivation of Ward identities etc.
For closed algebras,
one can invoke the formal argument that because of the invariance of the
measure for the classical fields $\phi^i$
the only possible Jacobian can
come from the ghost measure. It vanishes due to the fact that
$T^{\alpha}_{\beta \alpha} = 0$ for most of the interesting theories.
All such statements, however,
presuppose the introduction of a suitable regularisation scheme.
The choice of the scheme has a large influence on the actual form of
these contributions from
the measure. We will not discuss this issue, but
only restrict ourselves to the classical master equation.
For a detailed treatment of the use of Pauli--Villars regularisation for
studying BV, we refer to \cite{Troost}. Care should be taken when
following the formal arguments,
which was exemplified e.g. in \cite{Mijzelf}.

By integrating over the Nakanishi--Lautrup field $B_{\bar{a}}$, the
collective fields are fixed to zero and disappear upon integration.
Finally, integrating over the ghosts $c^{\bar{a}}$ of the shift symmetry
the conventional gauge fixing delta-function of BV is recovered:
\beq \delta \left( \frac{\delta^r \Psi}{\delta Y^{\bar{a}}} + Y^*
_{\bar{a}} \right). \eeq

\section{Open algebras}

By definition, one speaks of an {\it open algebra}
when the field equations of
the original action $S_0$ are needed to obtain the usual structure that
the
commutator of two gauge transformations is a linear combination of gauge
transformations:
\beq
\frac{\delta^r R^i_{\alpha}}{\delta \phi^j} R^j_{\beta} - (-1)^
{\epsilon_{\alpha} \epsilon_{\beta}} \frac{\delta^r R^i_{\beta}}
{\delta \phi^j} R^j_{\alpha} = 2 R^i_{\gamma} T^{\gamma}_{\alpha
\beta} (-1)^{\epsilon_{\alpha}+1} - 4 \frac{\delta^r S_{0}}
{\delta {\phi^j}} E^{ji}_{\alpha \beta} (-1)^i (-1)^{\epsilon_
{\alpha} + 1}     \label{algebra} \eeq
We will consider a set of gauge generators that is irreducible.
For open algebras, however, the na\"{\i}ve BRST operator (\ref{deltaN})
fails
to be nilpotent off-shell, owing to the term proportional to
field equations in (\ref{algebra}). This prevents us from following
the usual gauge-fixing procedure.
Nilpotency on all the fields
only holds when imposing the field equations of $S_0$.
Open algebras were first encountered in the study of supergravity
theories \cite{sugra}, and it proved possible in these cases
to introduce extra auxiliary fields to construct an
off-shell nilpotent BRST operator. After integrating out these
auxiliary fields, one ended up with terms quartic in the ghosts,
which one would not expect when applying straightforwardly the
Faddeev--Popov procedure. As the existence of these auxiliary fields
cannot be guaranteed in general, the quantisation procedure based
on their existence may be felt as unsatisfactory.

In \cite{open}, de Wit and van Holten proposed a general method for
quantising open algebras, without the need of having auxiliary
fields at one's disposal. Their basic observation is the following.
In order to derive, for example, Ward identities, one needs an action
$ S_{gf} $, which is invariant under some global transformation
reflecting the original gauge invariances, the BRST transformation
$\delta$. Usually, this is achieved by constructing $\delta$ such
that \begin{eqnarray} \delta S_0 & = & 0 \nonumber \\
          & \mbox{ and } &  \nonumber \\
   \delta^2 & = & 0 .\label{naivebrst}
\end{eqnarray} It is then clear that any $S_{gf} = S_0 + \delta X$
satisfies $\delta S_{gf} = 0 $, and that BRST invariant observables
will be independent of $X$. Both requirements (\ref{naivebrst})
can be dropped however, if one can just define an operator $\delta$
and a gauge-fixed action $S_{gf}$ such that $\delta S_{gf} = 0 $,
but where the former is not necessarily nilpotent and the
latter need not be decomposable in $S_0 + \delta X$. This still
allows for the derivation of the fundamental property $\langle \delta Y
\rangle = 0$, where $\langle Z \rangle$ stands for the vacuum
expectation value of an arbitrary operator $Z$.

The authors of \cite{open} succeeded in constructing such a $\delta$
and $S_{gf}$ for open algebras, generalising the results of
\cite{Kallosh} for supergravity. Consider the fermionic function
$F = \bar{c}_{\alpha} F^{\alpha} (\phi)$, where
the functions $F^{\alpha}$ are the gauge conditions. We will refer
to $F$ as the gauge fermion. Define
$ F_i = \frac{\delta F}{\delta \phi^i} $. Then $S_{gf}$ is
expanded as a power series in these $F_i$, where the linear term is
given by the Faddeev--Popov quadratic ghost action. The BRST
transformation that leaves this $S_{gf}$ invariant is obtained by
again adding an expansion in $F^i$ to the na\"{\i}ve transformation
laws. All coefficients in these expansions
are determined by demanding $\delta S_{gf}
=0 $, order by order in the $F_i$.

We will now show how the combination of this procedure with the
demand that the symmetry algebra includes the Schwinger--Dyson
BRST symmetry, leads straightforwardly to the antifield formalism
of Batalin and Vilkovisky for open algebras. The most important
aspect is the appearance of terms of order higher than one in the
antifields, which of course corresponds to the expansion in
powers of $F_i$ mentioned above.

We start by repeating the steps of the closed algebra case.
That is, we construct the na\"{\i}ve BRST operator (\ref{deltaN}) and
extend it to include the shift symmetry, leading to
(\ref{deltaNshift}).
Notice that the introduction of collective fields in no way
influences the answer to the question whether
the gauge algebra is open or closed,
as can be seen from calculating $\delta^2_N ( Y^{\bar{a}} -
{\cal Y}^{\bar{a}} ) $.

Consider now
\beq F = Y^*_{\bar{a}} {\cal Y}^{\bar{a}} + \Psi ( Y ) \eeq
as gauge fermion. The first term is clearly such that it fixes
the collective field to zero, while the second term is there
to fix the original gauge symmetry. This choice for $F$ gives
\beq \frac{\delta^r F}{\delta {\cal Y}^{\bar{a}}} = Y^*_
{\bar{a}} \eeq and \beq \frac{\delta^r F}{\delta Y^{\bar{a}}}
= \frac{\delta \Psi(Y)}{\delta Y^{\bar{a}}} \eeq as the quantities
$F_i$ in which we have to expand $S_{gf}$ and the BRST
transformations.

Following de Wit and van Holten \cite{open}, we then conjecture that
quantities
$ M_n^{\bar{a}_1 \ldots \bar{a}_n} $ exist, with the properties
\beq M_n^{\bar{a}_1 \ldots \bar{a}_i \bar{a}_{i+1} \ldots
\bar{a}_n} = (-1)^{(\bar{a}_i + 1)(\bar{a}_{i+1}+1)} M_n^
{\bar{a}_1 \ldots \bar{a}_{i+1} \bar{a}_i \ldots \bar{a}_n}
\label{antiM} \eeq
and
\beq \epsilon(M_n^{\bar{a}_1 \ldots \bar{a}_n}) = \sum_i \left(
  \epsilon_{\bar{a}_i} + 1 \right),\eeq
such that
\begin{eqnarray}   \label{Sgf}
   S_{gf} & = & S_0 (\phi - \varphi ) + (-1)^{\bar{a}} B_{\bar{a}}
   {\cal Y}^{\bar{a}} \nonumber \\
     &  & +  Y^*_{\bar{a}} c^{\bar{a}} - Y^*_{\bar{a}} {\cal R}
    ^{\bar{a}} ( Y - {\cal Y}) + \sum_{n \geq 2} \frac{1}{n}
      Y^*_{\bar{a}_1} \ldots Y^*_{\bar{a}_n} M_n^{\bar{a}_1
     \ldots \bar{a}_n} ( Y - {\cal Y} ) \\
    &  & +  \frac{\delta^r \Psi(Y)}{\delta Y^{\bar{a}}} c^{\bar{a}}
  \nonumber
\end{eqnarray}
is invariant under the BRST transformations
\begin{eqnarray}    \label{fullBRST}
   \delta Y^{\bar{a}} & = & c^{\bar{a}} \nonumber \\
   \delta {\cal Y}^{\bar{a}} & = & c^{\bar{a}} - {\cal R}^
   {\bar{a}} ( Y - {\cal Y}) + \sum_{n \geq 2} Y^*_{\bar{a}_2}
   \ldots Y^*_{\bar{a}_n} M_n^{\bar{a} \bar{a}_2 \ldots \bar{a}_n}
   (Y - {\cal Y}) \nonumber \\
   \delta c^{\bar{a}} & = & 0 \\
   \delta Y^*_{\bar{a}} & = & B_{\bar{a}} \nonumber \\
   \delta B_{\bar{a}} & = & 0 \nonumber .
\end{eqnarray}

The factor $\frac{1}{n}$ was introduced in (\ref{Sgf}) in order to
make all $B$-dependent terms cancel each other in $\delta S_{gf}$, as
\beq
\delta \left[ \frac{1}{n} Y^*_{\bar{a}_1} \ldots Y^*_{\bar{a}_n}
 M_n^{\bar{a}_1 \ldots \bar{a}_n} \right]
 = \frac{1}{n} Y^*_{\bar{a}_1}
\ldots Y^*_{\bar{a}_n} \delta M_n^{\bar{a}_1 \ldots \bar{a}_n}
+ (-1)^{\bar{a}_1+1} B_{\bar{a}_1} Y^*_{\bar{a}_2} \ldots
Y^*_{\bar{a}_n} M_n^{\bar{a}_1 \ldots \bar{a}_n} \eeq
owing to the permutation property (\ref{antiM}) of the $M_n$.
The term in $\delta S_{gf}$ that depends on $\Psi$
vanishes trivially, so that no non-linear terms in $\frac{\delta^r \Psi}
{\delta Y}$ are needed. In fact, this shows that our procedure is
independent of the choice of $\Psi$. This is a consequence of the
fact that the set of transformation rules (\ref{fullBRST}) is
nilpotent on $Y^{\bar{a}}$ and $c^{\bar{a}}$.
Taking these two facts into account, the
condition $\delta S_{gf}= 0$ leads to the following conditions,
obtained by equating order by order in the antifields $Y^*$ to
zero :
  \addtocounter{equation}{1}
$$
\begin{array}{lclcl}
   (Y^*)^0 & : & \frac{\delta^r S_0(Y-{\cal Y})}{\delta Y^{\bar{a}}}
    {\cal R}^{\bar{a}}(Y- {\cal Y}) & = &  0
             \mbox{ \hspace*{1.7cm} (\arabic{equation})} \\
    \addtocounter{equation}{1}
   (Y^*)^1 & : & \frac{\delta^r {\cal R}^{\bar{a}} (Y - {\cal Y})}
   {\delta Y^{\bar{b}}} {\cal R}^{\bar{b}} (Y-{\cal Y}) +
   (-1)^{(\bar{a} + 1)\bar{b}} \frac{\delta^r S_0( Y- {\cal Y})}
   {\delta Y^{\bar{b}}} M_2^{\bar{b} \bar{a}} ( Y - {\cal Y})
   & = &  0  \mbox{ \hspace*{1.7cm}  (\arabic{equation})}   \\
   (Y^*)^2 & : & \ldots \nonumber \\
   \ldots   \nonumber
\end{array}               $$
In principle, this gives equations at each order in the antifields,
which allow the construction of the $M_n$. Let us only study the two
above relations.
The term independent of the antifields
expresses the invariance of the classical action.
Considering the contribution to $\delta S_{gf}$ linear in the
antifields leads to two conditions, obtained by taking for the
$\bar{a}$-index $\phi^i$ and $c^{\alpha}$. In both cases $Y^{\bar{b}}$
runs over $\phi^j$ and $c^{\beta}$, so we get :
\begin{eqnarray}
   0 & = & \frac{\delta^r R^i_{\alpha} c^{\alpha}}{\delta \phi^j}
       R^j_{\beta} c^{\beta} + R^i_{\alpha} T^{\alpha}_{\beta \gamma}
     c^{\gamma} c^{\beta} + (-1)^{(i+1)j} \frac{\delta^r S_0}
    {\delta \phi^j} M^{ji}_2    \label{galgebra} \\
  0 & = & \frac{\delta^r T^{\alpha}_{\beta \gamma} c^{\gamma} c^{\beta}}
     {\delta \phi^j} R^j_{\mu} c^{\mu} + 2 T^{\alpha}_{\mu \nu} c^{\nu}
    T^{\mu}_{\beta \gamma} c^{\gamma} c^{\beta} + (-1)^{\alpha j}
    \frac{\delta^r S_0}{\delta \phi^j} M^{j \alpha}_2 .  \label{jacobi}
\end{eqnarray}
{}From (\ref{algebra}) and (\ref{galgebra}) it follows that
\beq
     M^{ij}_2 = 2 E^{ji}_{\alpha \beta} c^{\beta} c^{\alpha}.\eeq
We thus see that the coefficient of the term quadratic in the antifields
of the fields $\phi^i$ is completely determined by the non-closure
functions $E^{ji}_{\alpha \beta}$ of the algebra.
Furthermore, we see from (\ref{jacobi}) that
\beq M^{j \alpha}_2 = D^{j \alpha}_{\mu \nu \sigma} c^{\sigma} c^{\nu}
c^{\mu}. \eeq
By taking $\bar{c}_{\alpha}$ and $b_{\alpha}$ for $Y^{\bar{a}}$, we
find that the corresponding $M_2^{j \bar{c}_{\alpha}}$ and
$M_2^{j b_{\alpha}}$ are zero. This was to be expected, as they are
introduced as trivial pairs, decoupled from the original gauge algebra.
Let us now define
$ S_{BV} $ and $ S_{AD} $  by writing
\begin{eqnarray}
 S_{gf} (Y - {\cal Y}) & = & S_{BV} (Y- {\cal Y},Y^*) + (-1)^{\bar{a}}
B_{\bar{a}} {\cal Y}^{\bar{a}} + Y^*_{\bar{a}} c^{\bar{a}}
+ \frac{\delta^r \Psi}{\delta Y^{\bar{a}}} c^{\bar{a}}  \label{BVSol} \\
 & = & S_{AD}(Y -{\cal Y}, Y^*) +\frac{\delta^r \Psi}{\delta Y^{\bar{a}}}
         c^{\bar{a}}.
\end{eqnarray}
It is then clear that we recover the familiar form \cite{reducibleBV}
for $S_{BV}$ :
\beq
S_{BV} = S_0(\phi) - Y^*_{\bar{a}} {\cal R}^{\bar{a}} (Y)
  + \phi^*_i \phi^*_j E^{ji}_{\alpha \beta} c^{\beta} c^{\alpha} +
   \frac{1}{2} \phi^*_i c^*_{\alpha} D^{j \alpha}_{\mu \nu \sigma}
   c^{\sigma} c^{\nu} c^{\mu} + \ldots,\eeq
where the $\ldots$ stand for possible terms of more than quadratic
order in the antifields. Notice that the terms non-linear in the
antifields always contain at least one antifield $\phi_i^*$, because
$S_0$ only depends on the fields $\phi^i$.

We now remark that we can again decompose $S_{gf} = S_{AD} + \delta \Psi
(Y)$,
where $\delta S_{AD} = 0 $ and because $\delta^2 Y^{\bar{a}} = 0$.
Notice also that
\beq \delta {\cal Y}^{\bar{a}} = c^{\bar{a}} + \frac{\delta^l S_{BV}}
{\delta Y^*_{\bar{a}}} = \frac{\delta^l S_{AD}}{\delta Y^*_{\bar{a}}}
.\eeq
Taking all this into account, it becomes trivial to see that
\beq 0 = \delta S_{AD} = - \frac{\delta^r S_{BV}}{\delta
Y^{\bar{a}}} \cdot \frac{\delta^l S_{BV}}{\delta Y^*_{\bar{a}}}.
\eeq
So also in the case of open algebras, we recover the classical master
equation of Batalin and Vilkovisky. The gauge-fixing prescription is
again recovered. Notice that the extra terms in the BRST transformation
rules invalidate even the formal arguments on the absence of possible
Jacobians. Again, we will not discuss this important topic and refer to
\cite{Troost}.

Introducing collective fields after all
leads to the existence of a nilpotent BRST operator. In \cite{us} it
was shown that integrating out $B_{\bar{a}}$,$\,{\cal Y}^{\bar{a}}$ and
$c^{\bar{a}}$ leads to the so-called {\it quantum BRST operator}
$ \sigma X = (X,S_{BV}) - i \hbar \Delta X $, acting on quantities
$X(Y,Y^*)$.
This quantum BRST operator is nilpotent
since $S_{BV}$ satisfies the master equation and because of
properties of $\Delta$. This comes as no surprise, as the BRST operator
(\ref{fullBRST}) is nilpotent on $Y$,$\,Y^*$,$\,c$ and $B$.
At the classical level, this gives as BRST
operator the antibracket with $S_{BV}$, which is nilpotent. The
existence of a nilpotent BRST operator is of course a less trivial
result for open algebras.

Above we applied the procedure suggested by de Wit and van Holten
\cite{open} for the quantisation of open algebras. This leads us
to an explicit construction of $S_{BV}$ in (\ref{BVSol}), which is
then easily seen to satisfy the classical master equation of the BV
formalism. However, turning the argument around, one can see that
their procedure follows uniquely when starting from the requirement
of invariance of the gauge-fixed action under the Schwinger--Dyson BRST
symmetry. After integrating out the collective field, the latter
is given by
\begin{eqnarray}
        \delta Y^{\bar{a}} & = & c^{\bar{a}} \nonumber \\
        \delta c^{\bar{a}} & = & 0 \\
  \delta Y^*_{\bar{a}} & = &  \frac{\delta^l S_{gf}}{\delta Y^{\bar{a}}}
   .  \nonumber
\end{eqnarray}
{}From the study of the closed, irreducible gauge algebras
the form of $S_{gf}$ is then generalized to be always
\beq  S_{gf} = S_{BV}(Y,Y^*) + Y^*_{\bar{a}} c^{\bar{a}} +
    \frac{\delta^r \Psi}{\delta Y^{\bar{a}}} c^{\bar{a}}.\eeq
The requirement $\delta  S_{gf} = 0 $ then leads immediately, as
shown above, to the classical master equation for $S_{BV}$.
Together with the boundary condition that the term of $S_{BV}$
linear in the antifields acts as a source for the na\"{\i}ve BRST
transformations, leads uniquely to the de Wit--van Holten
quantisation for open algebras.

Now that we have a collective-field formalism at our disposal for
the derivation of the BV formalism for the quantisation of gauge
theories with open algebras, we can try to extend the results of
\cite{Mijzelf1} to include the
extended BRST-invariant quantisation of this kind of theories
\cite{BLT}. The natural generalization is then to allow for terms
proportional to an arbitrary power in the antifields
$Y^*_{\bar{a} a}$ \footnote{The extra index $a$ takes the values $1$,$2$.
It distinguishes the antifields associated with the BRST symmetry and
those associated with the anti-BRST symmetry, see \cite{Mijzelf1}.}
both in the transformation law for the collective field ${\cal Y}^
{\bar{a}}$ and in the gauge-fixed action. In the latter, also
terms with an arbitrary power in the collective field can be allowed.

In \cite{Mijzelf1}, it was found that the collective field formalism
leads to an unusual way of removing the antifield-dependent terms from
the path integral when one imposes extended BRST symmetry.
In the ordinary BV scheme and in \cite{BLT}, this is
done by just putting the antifields (including the collective field)
to zero after gauge-fixing. In contrast, in the
collective-field approach to extended
BRST symmetry, the collective field itself is fixed to
zero (this is at the heart of this approach), but the two antifields
$Y^*_{\bar{a} a}$ are removed by a Gaussian integration. This is
possible because the antifields only appear as linear source terms
for the extended BRST transformations.
We try to maintain this procedure for open algebras, i.e.
we try the decomposition
\footnote{The matrix $M_{\bar{a} \bar{b}}$ is needed for gauge-fixing
purposes, as is described in extenso in \cite{Mijzelf1}. Again, the
unbarred indices $a$,$\,b$ take the values $1$,$\,2$;
${\cal R}^{\bar{a}}_1$
denotes the BRST transformation of $Y^{\bar{a}}$, while ${\cal R}
^{\bar{a}}_2$ denotes its anti-BRST transformation.}
\beq S_{gf} = S_{BLT} + B_{\bar{a}} (-1)^{\bar{a}+1} M_{\bar{a} \bar{b}}
{\cal Y}^{\bar{b}} - \frac{1}{2}\epsilon^{a b} Y^*_{\bar{a} a} M_{\bar{a}
\bar{b}} Y^*_{\bar{b} b} -\frac{1}{2} \epsilon^{a b} {\cal R}^{\bar{a}}_a
M_{\bar{a} \bar{b}} {\cal R}^{\bar{b}}_b.\eeq
For closed algebras, the variation of the last term under a (anti-)BRST
transformation is cancelled by the variation of the term linear in the
collective field in $S_{BLT}$. The latter is the source term for the
composition of a BRST transformation with an anti-BRST transformation.
For open algebras this is no longer true, a term proportional to the
non-closure functions of the algebra appears in the variation of the
former. Also, owing to the non-linear terms, the gauge-fixing of the
$Y^*_{\bar{a} a}$ would also no longer be a Gaussian integral. Both
symptoms seem to indicate that still more terms need to be introduced,
even terms independent of antifields. One, however, has no guiding
principle when doing so. It thus seems very difficult to make contact
with \cite{BLT} in the case of an open gauge algebra.

\section{Reducible gauge algebras}

In \cite{reducibleBV}, Batalin and Vilkovisky gave the first
complete prescription for the quantisation of gauge theories
with arbitrary reducible gauge algebras.
Nevertheless, for the sake of completeness, we will also discuss
this case using the collective-field approach.
There are two aspects to the problem of quantising
reducible gauge theories. One is the construction of the BRST
operator and the ghost spectrum, and the other is the judicious
choice of the gauge fermion. We will only discuss the former,
the latter being extensively treated in \cite{reducibleBV}.
The collective field formalism has no bearing on the construction
of the gauge fermion, which is always considered to be available.

We start again from a classical action $S_0 (\phi)$, which has
gauge symmetries generated by $m$ operators $R^i_{\alpha}$. Suppose
now that $k$ quantities $Z^{\alpha}_{\bar{\alpha}}(\phi)$ exist,
such that
\beq R^i_{\alpha} Z^{\alpha}_{\bar{\alpha}} = 0. \label{redrel}
\eeq
This is just the expression that the original set of generators
was redundant, i.e. that not all $R^i_{\alpha}$ are independent. If the
$k$ $Z^{\alpha}_{\bar{\alpha}}$ are all independent, then one
speaks of a {\it first-stage-reducible} theory. Effectively, there
are then $m-k$ independent gauge symmetries.
The functions $Z^{\alpha}_{\bar{\alpha}}$ are not necessarily
independent,however, leading to second-stage-reducible theories
and so on. Here, we will only treat the first-stage theories
explicitly, higher-state theories allow for analogous constructions.

If one has not noticed the redundancy in the set of gauge generators,
the na\"{\i}vely gauge fixed action
\beq S_{gf} = S_0 + \bar{c}_{\alpha} \frac{\delta^r f^{\alpha}
(\phi)}{\delta \phi^i} R^i_{\alpha} c^{\alpha} + f^{\alpha}
b_{\alpha} \label{naiveSgf} \eeq
turns out to have a gauge symmetry in the ghost sector:
\beq \frac{\delta^r S_{gf}}{\delta c^{\alpha}} Z^{\alpha}_{\bar
{\alpha}} \epsilon^{\bar{\alpha}} = 0, \eeq
owing to the reducibility relations (\ref{redrel}).
Of course, this symmetry can be fixed by introducing so-called
ghosts for ghosts $\eta^{\bar{\alpha}}$, adding to the BRST
transformations of $c^{\alpha}$ a term $Z^{\alpha}_{\bar{\alpha}}
\eta^{\bar{\alpha}}$, as the functions $Z^{\alpha}_{\bar{\alpha}}$
are the gauge generators of this gauge symmetry in the ghost action.

The ghost action in (\ref{naiveSgf}) has however another gauge
symmetry, where the $\bar{c}_{\alpha}$ are the gauge fields. This
symmetry is generated by the left zero modes of $\frac{\delta f^{\alpha}}
{\delta \phi^i} R^i_{\alpha}$. These generators clearly depend on the
choice of the $f^{\alpha}$, and they do not enter the gauge algebra.
Their gauge fixing does not require more than trivial modifications
\footnote{By trivial modifications of the
BRST transformations we mean that one can always add pairs of fields $A$
and $B$, such that $\delta A = B$ and $\delta B = 0$, without changing
the physical content of the BRST cohomology. Such a pair of fields is
called a trivial system.} of the BRST transformations.

So, the situation can be summarized by saying that a nilpotent
BRST operator exists:
\begin{eqnarray}    \label{nilred}
      \delta \phi^i & = & R^i_{\alpha} c^{\alpha} \nonumber \\
      \delta c^{\alpha} & = & T^{\alpha}_{\beta \gamma} c^{\gamma}
        c^{\beta} + Z^{\alpha}_{\bar{\alpha}} \eta^{\bar{\alpha}}
         \\
      \delta \eta^{\bar{\alpha}} & = & A^{\bar{\alpha}}_{\mu \bar
         {\mu}}c^{\mu} \eta^{\bar{\mu}} + F^{\bar{\alpha}}
         _{\mu \nu \sigma} c^{\mu} c^{\nu} c^{\sigma}
          \nonumber     ,
\end{eqnarray}
supplemented with some trivial systems, introducing antighosts,
such that a suitable gauge fermion can be constructed, satisfying
the requirements of \cite{reducibleBV}. The functions $F$ and $A$
are determined from the nilpotency requirement of $\delta$ on
$c^{\alpha}$. Everything is then just the
same as for the case of irreducible, closed algebras, as far as
the collective field-formalism is concerned. Specifically,
an extended action linear in the antifields is obtained, where the
antifields act as sources for (\ref{nilred}).

However, even if the gauge algebra itself is closed, the BRST operator
(\ref{nilred}) may not necessarily be nilpotent off-shell, because terms
proportional to the field equations of $S_0$ can appear in the
reducibility relations (\ref{redrel}):
\beq R^i_{\alpha} Z^{\alpha}_{\bar{\alpha}} - 2 \frac{\delta^r
   S_0}{\delta \phi^j} B^{ji}_{\bar{\alpha}}
    (-1)^i = 0 .\eeq
The procedure to be followed is then the same as for open algebras,
as described above. For instance, (\ref{galgebra}) gets an extra term
$ R^i_{\alpha} Z^{\alpha}_{\bar{\alpha}} \eta^{\bar{\alpha}} $,
leading to an extra contribution $-2 B^{ji}_{\bar{\alpha}} \eta
^{\bar{\alpha}}$ to $M^{ij}_{2}$, so that finally a
term $- \phi^*_i \phi^*_j B^{ji}_{\bar{\alpha}} \eta^{\bar{\alpha}}$
appears in $S_{BV}$.
Other extra terms follow in exactly the same way from
(\ref{jacobi}).

\section{Conclusion}
We have thus shown that a collective-field derivation exists of the
BV quantisation scheme for theories with open and/or reducible
gauge algebras. The presence of terms proportional to field equations
in the original algebra and/or the reducibility relations necessitates
an approach like in \cite{open}. Demanding Schwinger--Dyson BRST, which
leads to the BV classical masterequation straightforwardly, leads to
this generalisation of classical BRST in a natural way.

\section*{Acknowledgements}
It is a real pleasure to thank P.H. Damgaard for many
encouraging discussions.

\end{document}